\documentclass[a4paper,conference]{IEEEtran}
\usepackage{amsmath}
\usepackage{url}
\newcommand{\precaption}{}
 
\newcommand{\postcaption}{ }

\usepackage{xcolor}

\usepackage{comment}

\usepackage{amssymb}

\usepackage{cleveref}

\usepackage{framed,url,caption,subcaption,graphicx}

\usepackage{amsmath,xcolor,soul, fontawesome,algpseudocode,algorithm,xspace,multirow,longtable,colortbl, lscape, pifont}
\usepackage{booktabs, siunitx,comment}

\definecolor{light-gray}{gray}{0.9}
\definecolor{darkgreen}{rgb}{0,0.5,0}
\definecolor{light-blue}{rgb}{0,.7,1}
\definecolor{red}{rgb}{.7, 0, 0}
\graphicspath{{./figures/} }
\crefformat{section}{\S#2#1#3}
\crefformat{subsection}{\S#2#1#3}
\crefformat{subsubsection}{\S#2#1#3}

\expandafter\def\expandafter\UrlBreaks\expandafter{\UrlBreaks
  \do\a\do\b\do\c\do\d\do\e\do\f\do\g\do\h\do\i\do\j%
  \do\k\do\l\do\m\do\n\do\o\do\p\do\q\do\r\do\s\do\t%
  \do\u\do\v\do\w\do\x\do\y\do\z\do\A\do\B\do\C\do\D%
  \do\E\do\F\do\G\do\H\do\I\do\J\do\K\do\L\do\M\do\N%
  \do\O\do\P\do\Q\do\R\do\S\do\T\do\U\do\V\do\W\do\X%
  \do\Y\do\Z}

\newcount\Comments  \Comments=1  
\newcommand{\kibitz}[2]{\ifnum\Comments=1\textcolor{#1}{#2}\fi}
\newcommand{\STOP}[1]  {\kibitz{red}   {[\textbf{STOP HERE} - compiled: \today]}}

\newcommand{\R}{\mathbb{R}}

\renewcommand{\precaption}{}

\renewcommand{\postcaption}{} 

\newcommand{\CERTBund}{Feed~1}
\newcommand{\CIRCL}{Feed~2}
\newcommand{\inThreat}{Feed~3}
\newcommand{\InnoTecSystem}{Feed~4}
\newcommand{\CthulhuSPRL}{Feed~5}
\newcommand{\DCSO}{Feed~6}
\newcommand{\Barncat}{Feed~7}
\newcommand{\Cryptolaemus}{Feed~8}
\newcommand{\Crimeware}{Feed~9}
\newcommand{\INCIBE}{Feed~10}

\usepackage[english]{babel}
\usepackage[utf8]{inputenc}
\usepackage{blindtext,comment}
   
\usepackage{ifthen}

 \usepackage{cite}
 
 \newcommand\blfootnote[1]{%
  \begingroup
  \renewcommand\thefootnote{}\footnote{#1}%
  \addtocounter{footnote}{-1}%
  \endgroup
}

\begin{document}

\newboolean{compact8pages}
\setboolean{compact8pages}{true} %

\title{Harnessing TI Feeds for Exploitation Detection}

\author{\IEEEauthorblockN{Kajal~Patel\IEEEauthorrefmark{1},        Zubair~Shafiq\IEEEauthorrefmark{1}, Mateus~Nogueira\IEEEauthorrefmark{4}, Daniel~Menasché\IEEEauthorrefmark{4}, Enrico~Lovat\IEEEauthorrefmark{2}, \\ Taimur~Kashif\IEEEauthorrefmark{1}, Ashton~Woiwood\IEEEauthorrefmark{3}, Matheus~Martins\IEEEauthorrefmark{2} }\IEEEauthorblockA{ \IEEEauthorrefmark{1}University of   California,  Davis, \IEEEauthorrefmark{4}Federal University of Rio de Janeiro,  \IEEEauthorrefmark{2}Siemens Corporation,  Princeton,  \IEEEauthorrefmark{3}University of   Iowa}}
 
\maketitle

\begin{abstract}
Many organizations rely on Threat Intelligence (TI) feeds to assess the risk associated with security threats. 
Due to the volume and heterogeneity of data, it is prohibitive to manually analyze the threat information available in different loosely structured TI feeds.
Thus, there is a need to develop automated methods to vet and extract actionable information from TI feeds. 
To this end, we present a machine learning pipeline to automatically detect vulnerability exploitation from TI feeds.  
We first model threat vocabulary in loosely structured TI feeds using state-of-the-art embedding techniques (Doc2Vec and BERT) and then use it to train a supervised machine learning classifier to detect exploitation of security vulnerabilities.
We use our approach to identify exploitation events in 191 different TI feeds. 
Our longitudinal evaluation shows that it is able to accurately identify exploitation events from TI feeds only using past data for training and even on TI feeds withheld from training. 
Our proposed approach is useful for a variety of downstream   tasks such as data-driven vulnerability risk assessment.
\end{abstract}

\section{Introduction}\label{sec:introduction}

Threat 
 Intelligence (TI) has become a cornerstone of collaborative data-driven security \cite{grandview}.
Many organizations rely on TI feeds to assess the risk associated with security vulnerabilities and deploy countermeasures in a timely manner. 
The interest in TI feeds has grown over the last few years, with an increasing amount and variety of TI feeds \cite{chismon2015threat}.
TI feeds are now available in many different flavors---from free to commercial, from many different providers ranging from third-party firms or public groups, and targeting various kinds of threats from phishing to botnets.\blfootnote{This paper appears at  IEEE International Conference on Cyber Security and Resilience (IEEE CSR), 2024.}

A key challenge in effectively leveraging TI feeds is coping with the volume of data. 
Large amounts of new threat information is regularly published, sometimes in excess of millions per day.
As the volume of threat information has increased, it has become infeasible to manually sift through them \cite{suciu2021expected}.
Thus, automated methods to systematically sift through TI feeds are needed. 
Another key challenge is that TI feeds are loosely structured due to heterogeneous data sources, software ecosystems, internal processes, maturity of the TI program, and staff expertise.
There has been progress over the last decade to standardize information sharing in TI feeds \cite{misp}.\footnote{The Malware Information Sharing Platform (MISP) \cite{misp} and the Structured Threat Information (STIX)   are two important initiatives in this direction. Both MISP and STIX establish formats that allow different organizations to share data in a uniform manner.}
While such standardization mitigates this challenge, a significant fraction of the information in TI feeds is made available within free-form text fields that are manually filled by security experts with differing styles.
These challenges make it non-trivial to automatically process the information in TI feeds to extract actionable information.

In this paper, we consider the problem of detecting whether information about the exploitation of vulnerabilities in the wild is present in a given TI feed event. 
Detection of vulnerability exploitation in the wild is crucial for risk assessment and incident response purposes. 
Consider, for instance, the Common Vulnerability Scoring System (CVSS) \cite{cvss}, one of the current de facto solutions to parametrize vulnerability risk. 
While CVSS does provide a subscore to track existence of exploits for vulnerabilities, it does not provide a subscore to track exploitation incidents \emph{in the wild}.  
The recently proposed Exploit Prediction Scoring System (EPSS) \cite{jacobs2021exploit} aims to bridge this gap by forecasting whether a vulnerability will be exploited following its public disclosure.
Thus, information about vulnerability exploitation in the wild can complement CVSS and EPSS based risk assessment strategies. 
To the best of our knowledge, there are currently no large-scale measurements leveraging TI feeds to assess exploitation in the wild, or tools to leverage those measurements   for risk assessment purposes.


The Known Exploited Vulnerabilities (KEV) catalog, maintained by CISA~\cite{CISAKEV},  is another important source for information on vulnerabilities that have been actively exploited in the wild, yet it faces significant limitations. Its static nature means it cannot adapt to the changing threat landscape, and its focus is  tailored to vulnerabilities that government agencies must address within certain deadlines, potentially missing broader threats.  
  Google Project Zero catalog also contains a list of vulnerabilities   exploited    in the wild, but faces similar limitations, e.g.,  the list currently contains  310 vulnerabilities~\cite{GoogleProjectZero}. Meanwhile, EPSS provides a dynamic prediction of exploit likelihood but lacks insight into the reasons behind its scores,  and is limited by its reliance on private data, which may not reflect wider environmental and organizational contexts~\cite{massacci2024holy}.

To address the aforementioned challenges, we propose a machine learning approach to automatically ingest TI feeds for detecting vulnerability exploitation in the wild and that can be tailored to specific organizational needs.
Our machine learning pipeline leverages natural language processing (NLP) based embedding techniques to parse and encode loosely structured information in TI feeds. 
To this end, we consider both static and dynamic embeddings techniques. 
Static embeddings generate the same embedding for the same vocabulary in different TI feed events: vector matrices which encode vocabulary into vectors are shared across events.  
Dynamic embeddings,  in contrast, aim at capturing vocabulary semantics in different contexts to address the issue of  context-dependent nature of information in TI feed events.
Specifically, we train a static embedding (Doc2Vec \cite{le2014distributed}) and dynamic embedding (BERT \cite{devlin2019bert})\footnote{We chose BERT over alternative models, such as ChatGPT, due to its free availability and fine-tunable weights, leaving comparisons with ChatGPT as subject for  future research.} in an unsupervised manner to capture the semantics of loosely structured information in different TI feeds. 
Moreover, we also propose purpose-built embeddings that are further trained (or fine-tuned) on data from TI feeds. 
TI2Vec and TIBERT represent specialized embeddings corresponding to Doc2Vec and BERT, respectively. 
%
%
We then leverage these embeddings along with labeled data to train a supervised classifier for detecting exploitation of security vulnerabilities from TI feeds. By training with TI data that accurately represents a given organization's   context, one can   mitigate biases, ensuring that   environmental nuances are   reflected.

\ifthenelse{\boolean{compact8pages}}{}{
Our dataset comprises events from  191 TI feeds, including data from the default feeds of MISP~\cite{misp}.\footnote{MISP is an Open Source Threat Information Sharing Platform, which comes pre-configured with some public data feeds.  In this work, we restrict our analysis to tlp:white (=public) events and anonymize the name of the source organizations. 
By focusing on anonymized tlp:white events we avoid restrictive non-disclosure agreements and biases and we allow for our results to be reproduced and extended by the scientific community.}
To assess the effectiveness of different classifiers in detecting exploitation events in the TI feeds, we perform temporal and spatial splits of our dataset.
In a temporal setting, the training and test sets contain past and future data, respectively. 
In a temporal+spatial setting, the training and test sets are setup such that the TI feed used for testing is set aside during training.  
While F1 scores vary over different time horizons, they average around 56--78\%. 
Overall, we find that dynamic BERT embeddings outperform static Doc2Vec embedding and purpose-built specialized embeddings (i.e., TIBERT, TI2Vec) outperform their pre-trained counterparts. 
%

Our machine learning approach to detect exploitation events from TI feeds is useful for a variety of downstream security tasks. 
For example, it can help organizations implement risk-aware patch management strategies. 
In particular, the exploitation events detected by our approach from TI feeds can be used to determine exploit code maturity level {impacting CVSS/EPSS scores.} 
Overall, our work suggests the feasibility of using TI feeds to automatically detect vulnerability exploitation in the wild, paving the way towards data-driven enrichment of vulnerability risk assessment. 

}

\begin{table}[!t]
    \centering \vspace{0.05in}
       \caption{Top-10 TI feeds ranked based on IoC count}
        {
        \resizebox{0.49\textwidth}{!}{
            \begin{tabular}{|l|r|r|r|r|c|l|}
            \hline
            Provider & Event count &  IoC count & Average number  & Average event~\    & Dominant IoC category (\%)  \\ 
            & &   ~\ ~\  & of IoCs per event & timespan (days)  &  \\
            \hline
            \hline 
            {\CERTBund}
            & 644 & 276,481 & 429.3 & 1.77 & network activity (69.65\%)  \\
            \hline 
            {\CIRCL} 
            & 1,234 & 165,292 & 133.9 & 45.86 & network activity (43.35\%)  \\
            \hline 
            {\inThreat} 
            & 130 & 139,852 & 1,075.8 & 21.29 & network activity (84.43\%)  \\
            \hline 
            {\InnoTecSystem} 
            & 6,955 & 52,803 & 7.6 & 0.72 & network activity (98.51\%)  \\
            \hline 
            {\CthulhuSPRL}   
            & 120 & 48,952 & 407.9 & 90.62 & network activity (84.73\%)  \\
            \hline 
            {\DCSO} 
            & 230 & 27,376 & 119.0 & 14.12 & network activity (56.62\%)   \\
            \hline 
            {\Barncat}  
            & 1,623 & 16,056 & 9.9 & 0.09 & external analysis (100\%)  \\
            \hline 
            {\Cryptolaemus} 
            & 613 & 15,953 & 26.0 & 0.00 & network activity (82.94\%)  \\
            \hline 
            {\Crimeware} 
            & 138 & 14,028 & 101.7 & 99.85 & payload delivery (81.10\%) \\
            \hline 
            {\INCIBE} 
            & 55 & 11,448 & 208.1 & 49.62 & payload delivery (68.52\%)  \\
            \hline
            {Other feeds} 
            & 2256 & 98,305 & 6.6 & 4.32 & network activity (45.98\%)  \\
            \hline
        \end{tabular}
        }
        }
        \label{table: statistics}
    \end{table}

Our key contributions and findings are summarized below:

 \textbf{Longitudinal analysis.} We conduct a longitudinal analysis of TI feeds, identifying the prevalence of   categories of events, such as network activity and payload delivery, across multiple organizations.  We also analyze the flow of information across TI feeds, identifying that some feeds play the role of sources whereas others behave as aggregators (Section~\ref{sec: data}).
    
\textbf{Exploitation identification. } We curate and label the events in TI feeds, e.g., indicating which of those correspond to exploitation as opposed to patch disclosures or security advisories.   To that aim, we mine for association  rules between event tags and exploitation.  Whereas tags are instrumental for curating the dataset, they require manual intervention, motivating fully automated  word embeddings (Section~\ref{sec:methods}).
    
\textbf{TI2Vec and TIBERT. } We capture semantics from loosely structured information in TI feeds using static and dynamic embeddings {(Doc2Vec and BERT, respectively and their specialized counterparts TI2Vec and TIBERT)} that are then used to train  classifiers to automatically detect exploitation of   vulnerabilities.  The training of the classifiers conforms to temporal   constraints, allowing us to assess the relevance of history on the accuracy of future predictions.   We also evaluate knowledge transfer across feeds, noting that such transfer is feasible  from TI feeds that   behave as sources to  aggregators, particularly when  the envisioned flow of knowledge is aligned with the  flow of information across TI feeds (Section~\ref{sec:results}).

\section{TI Feed Characterization}
\label{sec: data}
In this section, we analyze TI feeds in our dataset with respect to IoC categories as well as their temporal and spatial characteristics. Our findings indicate that automated methods are needed to efficiently interpret loosely-structured threat data for various downstream security tasks. 

\vspace{.05in} \noindent \textbf{Preliminaries.} 
A \emph{TI feed} is a general term used to describe a curated source of information such as IP addresses, file hashes, textual threat description, and other features providing context to existing and potential threats. 
All of these features are collectively referred to as \emph{Indicators of Compromise} (or IoCs).
Each TI feed, from a public or private provider,\footnote{In general, it is not always the case that there is a one-to-one correspondence between TI feeds and providers, as some TI feeds may contain events from multiple providers. In this work, however, as we anonymize providers (also known as organizations) we refer to feeds and providers interchangeably, with each provider corresponding to its own feed.} contains a list of \emph{events} and each event includes a collection of IoCs. 
%
An event could be a security incident such as a phishing attack, a botnet attack, or a security advisory regarding a software vulnerability.
Each event also includes timestamps indicating when the event was created and last modified.  
The IoCs within an event are also individually timestamped based on when they are created or last modified.  
The \emph{event timespan} is the difference in days between the latest and earliest IoC timestamps associated to that event. 
Each IoC is accompanied by a comment field which contains free-form textual description of the IoC as well as an IoC category. 
An event may also contain a comment field typically summarizing the security analyst's assessment of the event.
The list of IoCs, comments, and timestamps together constitute the \emph{attributes} of the event.
In addition to these attributes, an event may be assigned \emph{tags} by a security analyst. 
Tags follow standardized taxonomies.  
An event tagged as \emph{tlp:white}, for example,  contains information that can be shared with the general public according to the Traffic Light Protocol (tlp), whereas an event tagged as \emph{type:osint} contains information that is considered Open Source Intelligence (OSINT).

\begin{figure}[!t]
    \centering
     \includegraphics[width=\columnwidth]{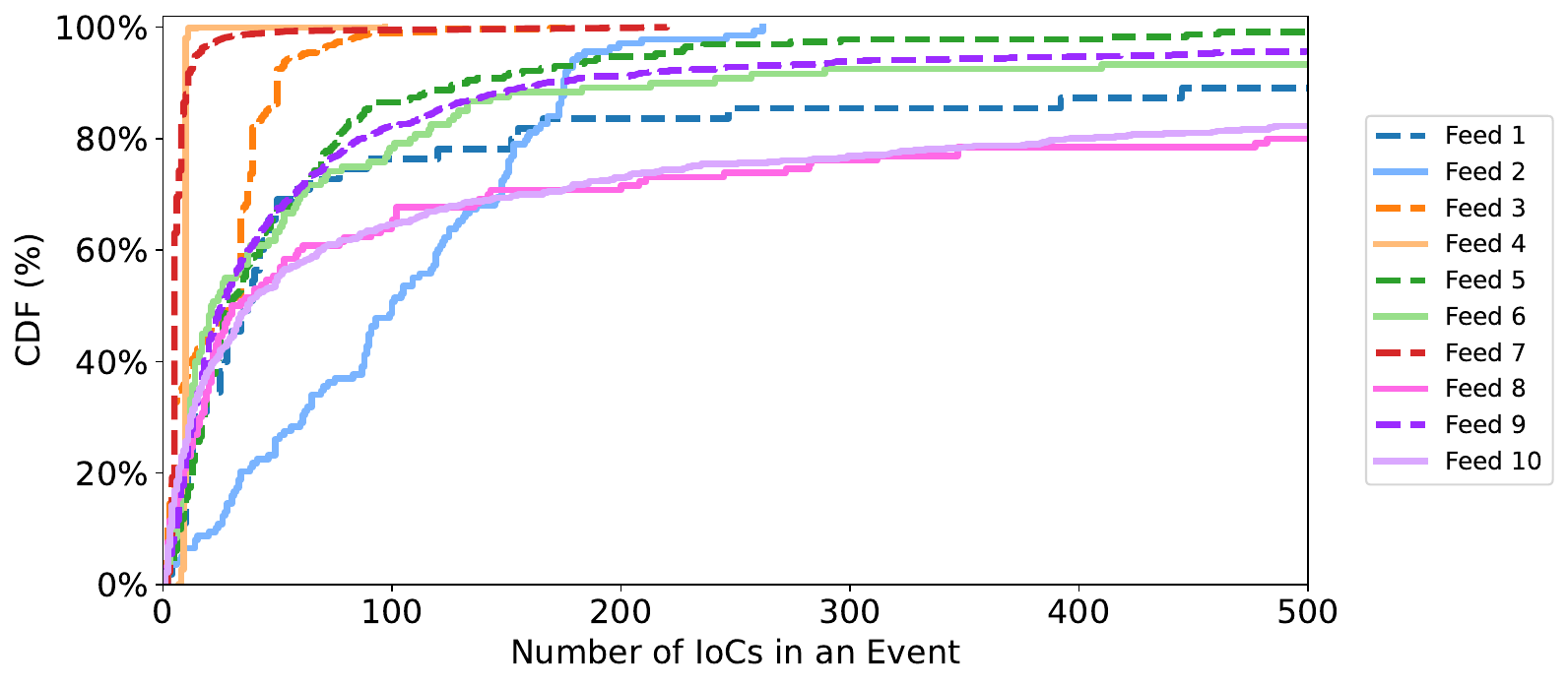}
    \caption{CDFs of   number of IoCs per event for   top 10 feeds.}
     \label{fig: iocs per event}
    \postcaption \vspace{-0.1in}
\end{figure}

\vspace{.05in} \noindent \textbf{Summary statistics.} 
Our dataset is comprised of  a collection of 191 TI feeds that come from a variety of public and private providers.
In total, our dataset contains about 14 thousand events and 866,541 IoCs over the duration of 7 years. 
Table \ref{table: statistics} lists key statistics of top-10 TI feeds ranked based on the number of IoCs.
The largest contributor is {\CERTBund}{} (a government organization) which contributes 276,481 IoC references with an average of 429 IoCs per event.
We note that the number of IoCs per event varies by several orders of magnitude across different feeds. 
Some feeds such as {\inThreat} have more than a thousand IoCs per event on average while others such as {\InnoTecSystem} have less than 10. 
Figure \ref{fig: iocs per event} further plots the distribution of the number of IoCs per event across different feeds. 
We note that even for feeds with a high average number of IoCs per event, the majority of events contain only a handful of IoCs.
For example, each of the top 10 feeds have at least 70\% of their events having no more than 200 IoCs each.
Further, all of the top 10 feeds, except {\Crimeware}, have at least 60\% of their events having no more than 100 IoCs each.

\subsection{IoC Categories}
Different TI feeds are geared towards different types of IoCs.
Specifically, each IoC is classified by the feed vendor into one of several IoC categories.\footnote{IoC categories are self-reported by each TI feed vendor. We do not expect different TI feeds to use exactly the same categorization procedure.}
Table \ref{table: statistics} shows the most dominant IoC categories for top-10 TI feeds in our data.
We note that a majority of popular TI feeds contain mostly network activity related IoCs while others contain IoCs related to payload delivery and external analysis. 
Network activity IoCs generally consist of IP addresses, hostnames, domains, and URLs.
Payload delivery IoCs generally consist of filenames, hashes (e.g., sha1, sha256 and md5), and malware samples.
External analysis IoCs generally consist of textual analysis by security experts and links to other pertinent organizations.

\begin{figure}[!t]
    \centering
     \includegraphics[width=0.9\columnwidth]{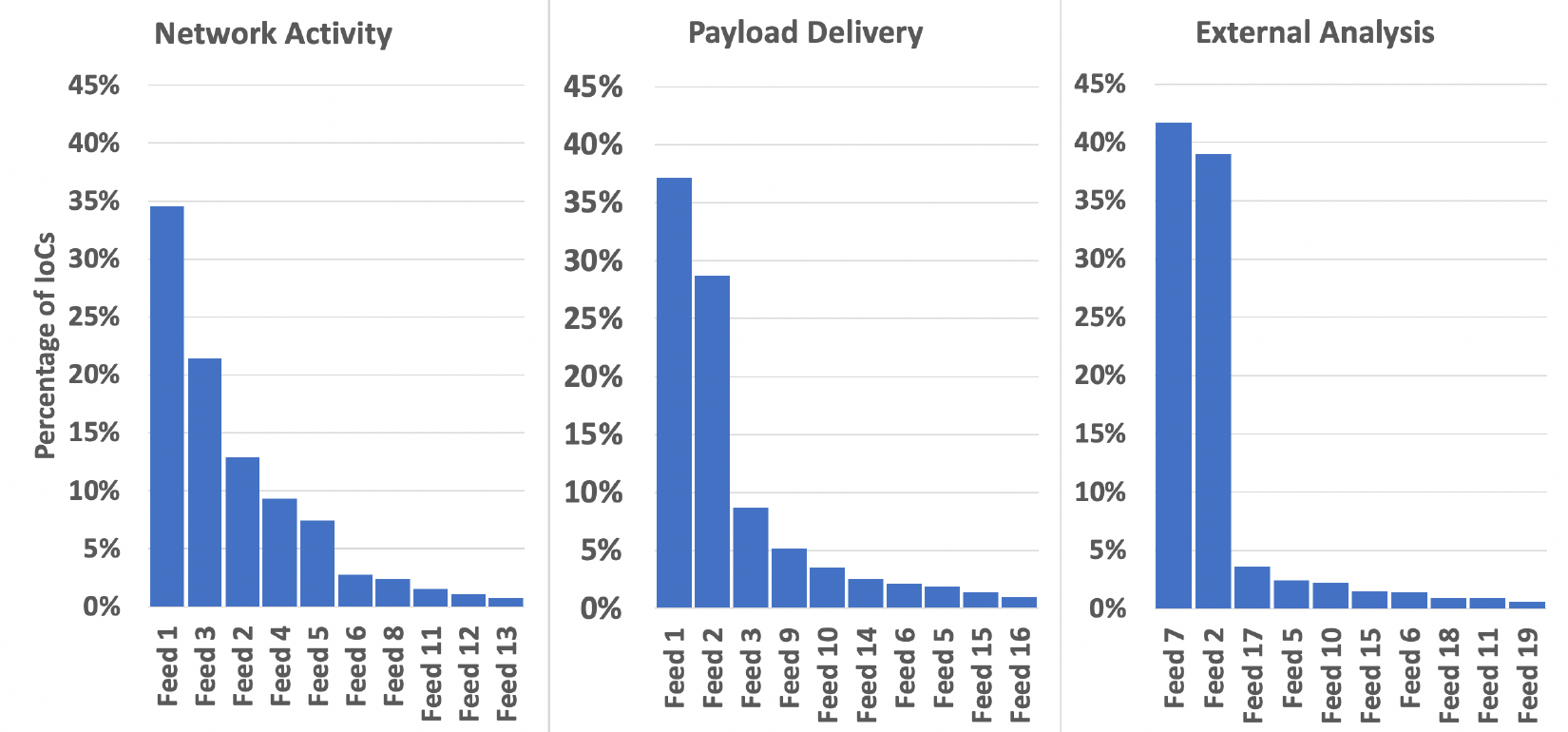}
    \caption{Contribution of  TI feeds to popular IoC categories.}
     \label{fig:FeatureCounts}
     \vspace{-0.25in}
    \postcaption
\end{figure}

Next, we analyze the relative contribution of different TI feeds across popular IoC categories. 
Figure~\ref{fig:FeatureCounts} plots the percentage of IoCs contributed by different TI feeds across popular IoC categories. 
We limit our analysis to top-3 IoC categories that comprise 87\% of the IoCs in our data.
Less popular IoC categories include artifacts dropped, social networks, support tools, etc.
The largest IoC category is {network activity}, which accounts for 557,532 IoCs.  
Across different TI feeds, {\CERTBund} and {\inThreat} together contribute more than 50\% of network activity IoCs. 
The second largest IoC category is {payload delivery}, which accounts for 220,321 IoCs. 
Across different TI feeds, {\CERTBund} and {\CIRCL} each contribute about 30\% of payload delivery IoCs. 
The third largest IoC category is external analysis, which accounts for 38,531 IoCs. 
Across different TI feeds, {\Barncat} and {\CIRCL} each contribute almost 40\% of external analysis IoCs.

Next, we analyze IoC  comments. 
We find that all except one feed have a majority of events with at least one IoC with a comment. 
Across these feeds, on average, 67.7\% of IoCs have a comment and the comment length is 27.7 characters. 
We identified three categories of comments: 
additional description of an attack, format of the IoC, or logistical upload information. 
The first type of comments usually contain a note/advice from a security expert.
For example, an IoC within one of our feeds 
 had an additional information comment: ``Kill switch domain. Monitor, do not block. (Source: Cisco Umbrella).'' 
Such comments could be useful in determining how to proceed when responding to a security incident. 
The second type of comments describe the format of IoCs. 
For example, an IoC within one of the considered feeds  
contained the comment ``RAT name,'' thus providing useful information that the IoC mentions the name of a specific remote access trojan. 
The third type of comments contain other logistical information that also provide context for the IoC. 
For example, an IoC 
contained the comment ``This attribute has been automatically imported.'' 
This information could be necessary when prioritizing which events or IoCs need to be analyzed first. 
Thus, we conclude that the information in these free-form textual comments provide deeper insights into security incidents. 

\newcommand*{\MinNumberTemporal}{0}%
\newcommand*{\MaxNumberTemporal}{216714}%

\newcommand{\ApplyGradient}[1]{%
        \pgfkeys{/pgf/fpu,/pgf/fpu/output format=fixed}%
        \pgfmathsetmacro{\PercentColor}{(#1 + 10000) / \MaxNumberTemporal * 100}%
        \edef\x{\noexpand\cellcolor{yellow!\PercentColor}}\x\textcolor{black}{#1}%
        \pgfkeys{/pgf/fpu = false}
}
\newcolumntype{R}{>{\collectcell\ApplyGradient}c<{\endcollectcell}}

\renewcommand*{\MinNumberTemporal}{0}%
\renewcommand*{\MaxNumberTemporal}{307963}%

\renewcommand{\ApplyGradient}[1]{%
        \pgfkeys{/pgf/fpu,/pgf/fpu/output format=fixed}%
        \pgfmathsetmacro{\PercentColor}{(#1) / \MaxNumberTemporal * 100}%
        \edef\x{\noexpand\cellcolor{yellow!\PercentColor}}\x\textcolor{black}{#1}%
        \pgfkeys{/pgf/fpu = false}
}

\ifthenelse{\boolean{compact8pages}}{}{
\subsection{Temporal analysis}
Since an event is essentially a timeseries of IoCs, we start our temporal analysis by breaking down event timespan across different TI feeds. Recall that the \emph{event timespan} refers to the interval between the lowest and the highest timestamp of the IoCs in the event.  Timestamps, in turn, are instrumental to determine for how long IoCs should be monitored~\cite{tostes2023learning}.

As shown in Table \ref{table: statistics}, average event timespan significantly varies across different TI feeds. 
Some TI feeds such as {\CERTBund} have shorter average event timespans of less than a couple of days while others such as {\CthulhuSPRL } have longer average event timespans of up to 3 months. 
Note that several TI feeds such as {\InnoTecSystem} and {\Barncat}  have the average event timespan of less than one day. 
In fact, the average timespan of {\Cryptolaemus} is zero, which means that all IoCs in each of its events have the same timestamp. 
Figure \ref{fig: event timespan} further plots the distribution of event timespan across different feeds. 
We note that 9 of the top 10 feeds have at least 70\% of their events having timespans of no more than 4 days.
{\Crimeware} is distinct in that its events tend to have much longer timespans, with under 30\% of its events having timespans of fewer than 30 days.  In Section~\ref{sec:longitudinaleval} we utilize temporal information to split the data into training and test sets. In Section~\ref{sec:results} we evaluate how the recency of posts influences their utility.

\begin{figure}[!t]
    \centering
     \includegraphics[width=\columnwidth]{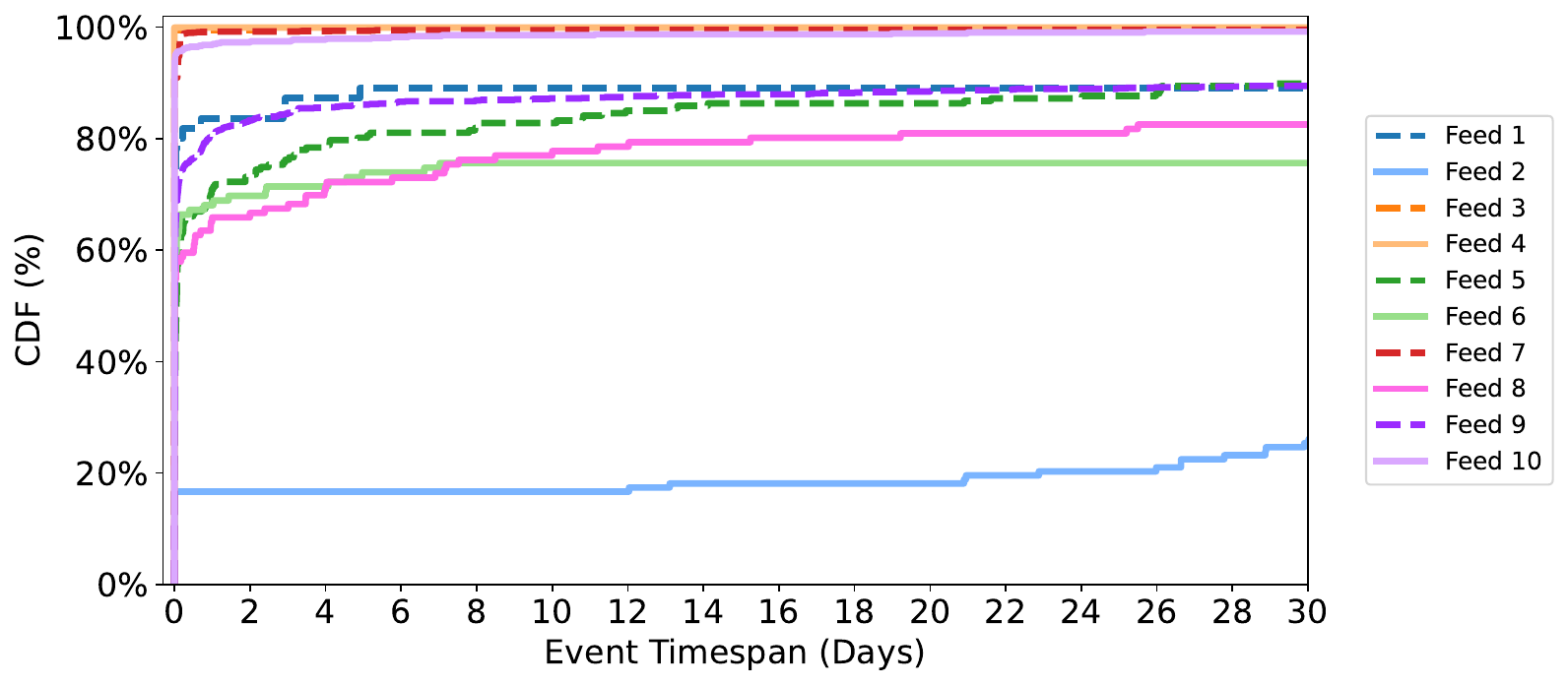}
    \caption{CDFs of the event timespan for the top 10 feeds.}
     \label{fig: event timespan} \vspace{-0.1in}
    \postcaption
\end{figure}
}

\subsection{Spatial Analysis}
Next, we analyze the IoC overlap to characterize information redundancy across TI feeds.  
Note that we restrict ourselves to IoCs containing specific IP addresses and hashes.
Some feeds share only a few IoCs with other feeds, while others share a lot of IoCs with several other feeds. 
For example, {\Cryptolaemus} shares less than 4 IoCs with all but one of the top-10 TI feeds. 
In contrast, many TI feeds tend to have a significant number of their IoCs in common with  particular sets of other feeds.
For instance, {\CIRCL} shares over 250 IoCs with  each of 6 of the other top feeds, with over 900 IoCs shared with {\DCSO} and over 1000 IoCs shared with {\INCIBE}.
Similarly, {\CERTBund} shares upwards of 200 IoCs with {\CIRCL} and {\inThreat} and over 1700 with {\DCSO}.
Such overlap between different TI feeds is not surprising and has been reported in prior work on IP blocklists \cite{li19tealeaves,Ramanathan20blacklists}.

\begin{figure}[t!]
\centering
\includegraphics[width=0.6\columnwidth]{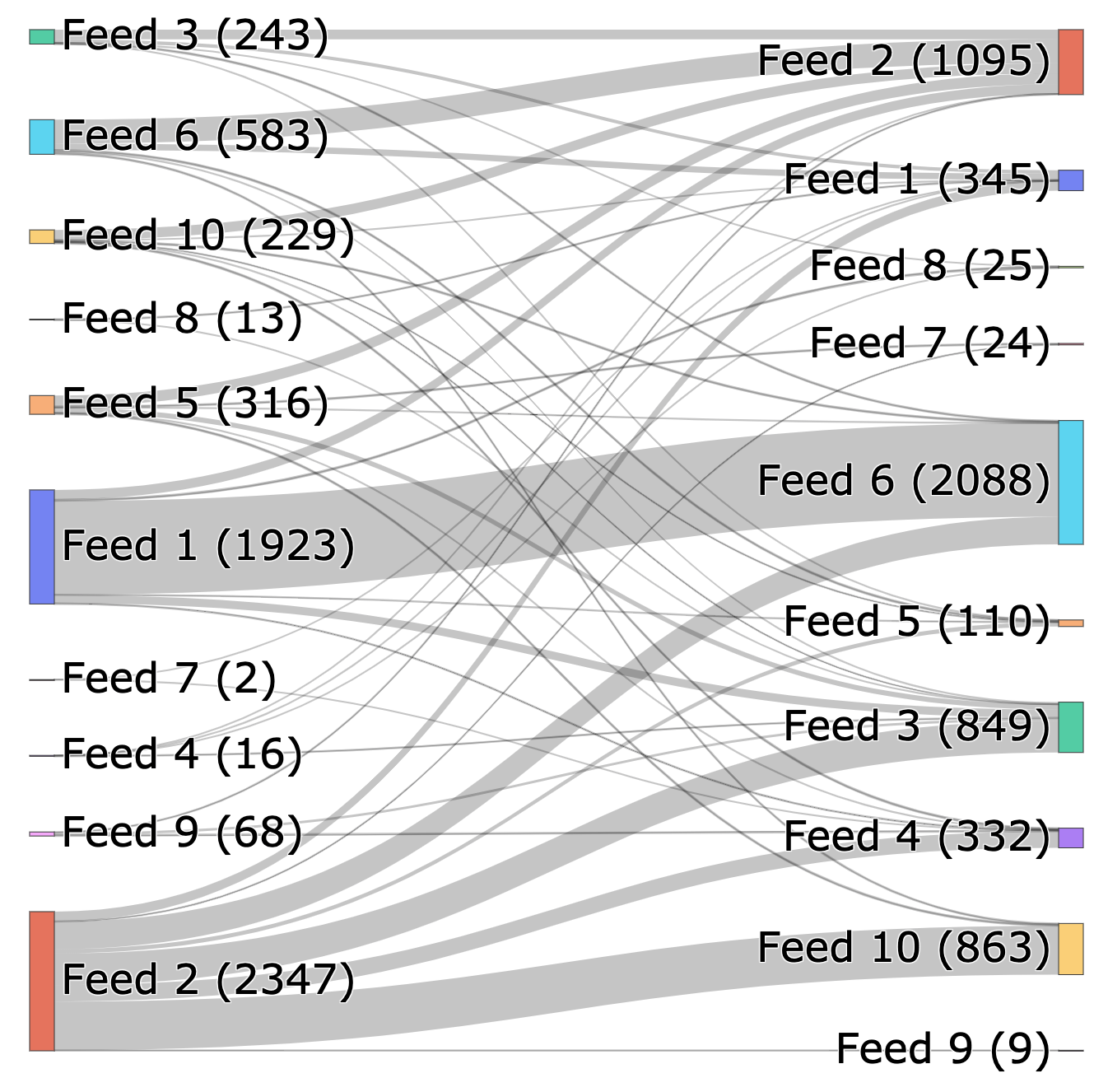}
    \caption{Flow of common IoCs across top-10 TI feeds. } 
    \label{figure:sankey}
    \vspace{-0.2in}
\end{figure}

We next analyze the flow (or temporal ordering) of common IoCs across TI feeds. 
Specifically, in contrast to the IoC overlap analysis above, we take into account which feed publishes a common IoC before the other.
Figure~\ref{figure:sankey} plots a Sankey diagram to visualize the flow of IoCs across the top-10 TI feeds in our dataset. 
First, we note that some TI feeds are mainly the source of common IoCs. 
For example, {\CERTBund} is the source of approximately 84.8\% of its common IoCs.
Second, some TI feeds are mainly the sink of common IoCs. 
For example, {\DCSO} is mainly the sink of IoCs from {\CERTBund} and {\CIRCL}. 
This shows that {\DCSO} basically aggregates a subset of IoCs from multiple TI feeds.
Third, other feeds are more symmetric in being the sources/sinks of common IoCs.
For example, {\CIRCL} is the source of 68\% of its common IoCs with other feeds.
Thus, we conclude that TI feeds may have source/sink relationships for common IoCs with other feeds.

\vspace{.05in} \noindent \textbf{Takeaways.} 
Our characterization highlighted three main insights. 
First, TI feeds contain various categories of information about different threats in a loosely-structured format. 
Second, the information in TI feeds and specific events is not static---it is dynamically updated and spread over often long time intervals. 
Third, TI feeds complement each other by providing differing amounts and types of IoCs but do sometimes contain redundant information. 
As we discuss next, there is a need to develop automated methods to analyze loosely-structured, spatio-temporal threat data in different TI feeds for downstream risk assessment, incident response, and patch management tasks. \ifthenelse{\boolean{compact8pages}}{For further details, we refer to~\cite{techrep}.}{}

\section{TI Feed Classification}
\label{sec:methods} 
The increasing volume of loosely-structured information in TI feeds renders them unwieldy for manual sifting, calling for methods to extract actionable information in an automated manner.    
In this section, we present a machine learning pipeline to automatically analyze and classify TI feed events.
Our proposed machine learning pipeline first leverages state-of-the-art natural language processing (NLP) techniques to extract semantic information from TI feeds and then trains supervised machine learning models for classification.

\subsection{Problem statement}
We focus on the problem of detecting exploitation of security vulnerabilities from TI feeds, {classifying  TI feed events for that matter}.\footnote{Note that this problem is related but different compared to vulnerability weaponization (i.e., whether an exploit for a particular vulnerability exists) \cite{suciu2021expected,Frei06vulnerabilityanalysis}. In particular, weaponization can be determined in a controlled environment, whereas exploitation requires measurements in the wild.} 
%
%
This is a useful undertaking because exploitation information is crucial for threat response,  vulnerability management and patch prioritization. 
The most concrete application is the recently proposed Exploit Prediction Scoring System (EPSS) \cite{jacobs2021exploit}, which specifically aims to predict the likelihood of vulnerability exploitation after disclosure. 
EPSS scores (i.e., likelihood that a vulnerability will be exploited within 3 months) are calculated based on a regression model that takes into account features such as the vendor's name and whether a proof-of-concept exploit has been developed~\cite{jacobs2021exploit}.   Whereas EPSS is parametrized using  private data from Fortinet and other partners, TI feeds provide open data that can be filtered  according to the organization needs,  allowing users to leverage  and tune public metrics and catalogs such as EPSS,  CISA KEV and Google Project Zero to their needs.

Measurement and detection of exploitation events in the wild essentially provides the ground truth data geared towards the context of a given organization. 
  Indeed, after an  event is  classified as an exploitation, the set of vulnerabilities in the attack surface of the environment where the exploitation occurred are natural candidates to have their CVSS/EPSS scores updated. 
%
%
%
%
%
%
%
%
%
Thus, we aim to   build models to automatically classify TI feed events into two classes: \textit{exploitation} and \textit{non-exploitation}.\footnote{\url{http://inthewild.io}, for instance, provides a  curated dataset with artifacts that are manually classified between those two classes. Our aim is to show the feasibility of automatic classification of TI feed events.}

\subsection{Ground truth}
We need labeled ground truth to train and test our machine learning classification pipeline.
Thus, we discuss our approach to bootstrap ground truth using a small set of TI feed events that are painstakingly labeled by experts manually.

\textbf{Manual labeling by experts. }
We begin by filtering TI feed events that mention specific vulnerabilities (i.e., CVE–ID) in their textual descriptions, comment, or IoC attributes. 
The filtered TI feed events are then manually labeled by two independent experts, with 10\% concurrently labeled events achieving the inter-rater agreement score of 89\%.
The experts used the following code book to manually label exploitation events.
Events that are labeled exploitation:   
(1) have a definite malicious components such as APTs, DDoS, and trojan attacks; 
(2) are part of a well-known malicious campaign; 
(3) mention a well-known hacker group/organization; or 
(4) contain keywords such as espionage or campaign (given suitable surrounding context).  
In contrast, events that are labeled non-exploitation:   
(1) are updates to existing software without a malicious component for that update;
(2) contain keywords such as patch, update, reset (given suitable surrounding context); 
(3) have no clear-cut indication of malicious action; or
(4) contain insufficient data or ambiguous terminology to be classified as exploitation. 
This manual procedure was used to label four thousand events.

\textbf{Event tags. }
\label{subsec: tags}
Each event in TI feeds may contain one or more tags.\footnote{Tags are referred to as `machine tags' in MISP.} 
Composed of a namespace, predicate, and optional value, a  tag of the form namespace:predicate=value can be utilized to further classify an event. ``misp-galaxy:threat-actor=Sofacy'' and ``tlp:white'' are examples of tags in an exploitation event. 
These tags are part of the MISP taxonomies~\cite{misp}, which allow various organizations to adopt standard classification vocabulary. 
The tags are manually assigned to each event and are designed for automated consumption by downstream tasks (e.g., intrusion detection systems).
There are drawbacks to this approach as bias from a security expert could sway which tags are being added and tagging each event is time intensive. 
On the other hand, it can be easier to control quality with manual tagging. 
Since every event in our dataset is annotated with at least one tag, tags could be utilized as a baseline for classification analysis.

\begin{figure}[!t]
    \centering
     \includegraphics[width=\columnwidth]{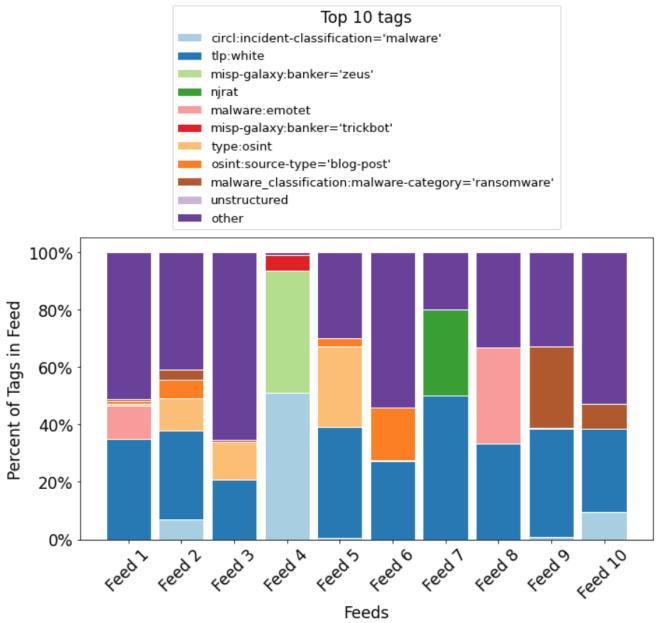}
       \precaption
          \vspace{-.2in}
    \caption{Breakdown of   common tags throughout top 10 feeds}
    \label{fig:FrequencyTagBreakdown}
    \postcaption
    \vspace{-.26in}
\end{figure}

We analyze the distribution of tags in the top-10 feeds in our dataset. 
Figure \ref{fig:FrequencyTagBreakdown} plots the percentage of each tag type in terms of all tags found per feed. 
%
%
In most feeds, the events typically contain 2-3 tags. 
Common tags across feeds include ``circl:incident-classification=malware'', ``malware:emotet, type:\-osint'', and ``malware{\_}classification:malware-category=ransomware''. %
The specific nature of each tag helps to verify the focus of an event. 
Since tags are manually added and hold pertinent information about an event, they may be useful in determining exploitation incidents.

\textbf{Automated labeling based on event tags. }
To expand ground truth beyond the four thousand manually labeled events, the rest of the labeling process was automated using these event tags. 
%
Specifically, we extracted association rules between event tags and exploitation on manually labeled events. 
We used the {Apriori} algorithm with support $\geq$ 0.001, confidence $\geq$ 0.5, and lift $\geq$ 1 to shortlist approximately 735 rules. 
Note that we filtered some unrelated tags (e.g., \texttt{tlp}) before extracting association rules. 
Once the rules which mapped combinations of tags to exploitation were generated, these rules were used to determine the ground truth for the remaining ten thousand unlabeled events. 
More specifically, an event was labeled as an \textit{exploitation} if and only if all of the tags in at least one of the shortlisted rules matched the tags of the event. 
If not, the event was marked as \textit{non-exploitation}. 
For example, the Apriori algorithm determined a rule \texttt{sectorfinancial} \& \texttt{kill-chain:delivery} \& \texttt{kill-chain:command and control} $\rightarrow$ \textbf{Exploitation}, meaning that if each of “sectorfinancial”, “kill-chain:delivery”, and “kill-chain:command and control” tags are present then it  would be labeled as an exploitation event. %

\textbf{Label distribution. }
We note that 11\% of  events in the top 10 feeds were labeled as exploitation and 
%
%
  the distribution of threat events varies across different feeds. 
For example, {\inThreat} and {\INCIBE}  had approximately 89\% and 87\% of their events labeled as exploitation, respectively.
This is in stark contrast to the ground truth labeling results of feeds such as {\InnoTecSystem}, {\Barncat}, and {\Cryptolaemus}, all of which had no exploitation events.
Across all TI feeds in our dataset, 82\% of the events were labeled as non-exploitation and the remaining 18\% were labeled as exploitation.

\textbf{Summary. }
The rule-mining mechanism   effectively automates the labeling procedure, but leverages  manually curated event tags for its purpose.  The classifiers  considered in the remainder of this paper, in contrast, are useful even when  event tags are not  present, e.g.,  for feeds with no tags, or as soon as IoCs are  produced.   To evaluate the accuracy of those classifiers, in turn,  it is key to conform to temporal and spatial constraints, as described next.

\subsection{Longitudinal evaluation} \label{sec:longitudinaleval}
A key challenge in detecting incidents such as vulnerability exploitation is their constantly evolving arms race nature. 
Thus, it is important to evaluate the accuracy of detection models in a longitudinal manner as classifiers need to evolve over time in response to the dynamic threats. 
To this end, we use two methods to split the dataset for training and testing: \emph{temporal} and \emph{temporal+spatial}.

\textbf{Temporal splits. }
To evaluate the accuracy of our classification models in a longitudinal setup, we split our entire dataset into distinct time windows such that the earlier windows are used for training and the later windows are used for testing. 
Thus, we make sure that the events in the train set occur prior to those in the test set.
We consider two different approaches to longitudinally split our dataset: aggregate window and equal window. 
%
%
Both aggregate window and equal window splits divide data into 11 time windows such that each window contains an equal number of events. 
In aggregate window split, the model uses the events in the next time window for testing and the events in all previous time windows for training. 
Thus, the amount of training data increases over time as more time windows are accumulated.
%
In equal window split, the model uses the events in the next time window for testing and the events in the immediately preceding time window for training. 
Thus, the amount of training data remains the same over time in equal window splits. 
Such temporal splits would help us evaluate whether the accuracy of our machine learning pipeline is sensitive to the amount and recency bias of training data.

\textbf{Temporal+spatial splits. }
We further build upon the aforementioned temporal splits by additionally splitting events across different TI feeds. 
Specifically, we consider the scenario where a feed is not available during the training but its events need to be classified at the test time. 
Thus, the test set is restricted to contain events only from one feed and the associated train set contains no events from that feed while respecting the aforementioned temporal splits. 
For example, consider  {\CIRCL}. 
If we split spatially, the train set of each window will not contain any events from {\CIRCL}, while the test set will only contain events from {\CIRCL}.
Such temporal+spatial splits would help us evaluate whether the accuracy of our machine learning pipeline generalizes to other previously unseen feeds.

\subsection{Features}
Next, we explain our feature extraction pipeline.

\textbf{Pre-processing. }
Before features are extracted, we conduct standard pre-processing, including splitting the text on all special characters, such as hyphens, and removal of stop words.
Some IoCs, such as hashes, might refer to a malware payload, and receive special treatment. 
As threats have polymorphic capabilities to mask file signatures, most hashes are unique to one feed.
This, in turn, implies that information about hashes cannot be readily generalizable across feeds.   
To address this issue, we replace all instances of hashes with an abstract representation noting the existence of a hash value.

\textbf{Feature extraction. }
We consider two different embedding-based feature extraction techniques. 
First, we use a context-independent static embedding technique, called Doc2Vec \cite{le2014distributed}.
Doc2Vec maps all words in an event to a feature vector such that it captures the relationship between a word given its neighboring words. 
Second, we use a content-dependent dynamic embedding technique, called Bidirectional Encoder Representations from Transformers, BERT~\cite{devlin2019bert}. 
BERT maps all words in an event to a feature vector using an attention-based model that learns contextual relations between words in a text. 
We use BERT base uncased  in~\cite{xiao2018bertservice} to encode our input data into fixed-length vector representations. 
As compared to Doc2Vec, which generates the same embedding for the same word in different contexts, BERT is better able to capture the semantics by modeling the surrounding context of an input word. 

For both Doc2Vec and BERT, we not only consider pre-trained models but also train purpose-built models that are further trained on data from TI feeds. 
For Doc2Vec, we train TI2Vec in an unsupervised manner using the corpus of 191 TI Feeds in our dataset. Training is done over 15 epochs. This customized embedding aims to capture the context of domain specific vocabulary that is typically found in TI feeds. With BERT for sequence classification, we train TIBERT in a supervised manner from the standard BERT-base-uncased model with our TI dataset. We utilize a classification layer to fine-tune it over 10 epochs.
%
%

%

\textbf{Feature selection. }
We perform feature selection to avoid over-fitting.
%
%
For {Doc2Vec}, we select  1000 features. 
For BERT, input is transformed into 768-dimensional vectors.

\textbf{Supervised classification. }
Given the features and labeled dataset, we train different supervised machine learning algorithms to classify TI feed events.  
We begin by considering Naive Bayes and Decision Tree classifiers, as both serve to classify  non-linearly separable classes, with Naive Bayes relying on conditional independence among features for that purpose.  
Following the discussion in Section~\ref{sec: data}, there are significant inter-dependencies across features within each class, suggesting that Decision Trees are better suited than Naive Bayes. 
As a natural extension to  Decision Trees, we also consider the AdaBoost ensemble classifier.  
Boosting in that context consists of using many  low-fidelity Decision Trees, which together are supposed to outperform a single tree. 
While the Decision Tree is amenable to interpretability, ensemble classifiers tend to provide better classification performance.
Finally, TIBERT has a built-in head that classifies the transformer's vector representation of the input based on what was learned during training, i.e., different from the other considered approaches, TIBERT learns to produce its embeddings and to classify events through a unified framework.

\section{Experimental Evaluation}
\label{sec:results}

To evaluate the accuracy of our classifiers, we use three well-known metrics: precision, recall, and F1 score. 
Precision is the fraction of presumed exploitation events that are correctly detected.
Recall is the fraction of exploitation events that are detected. 
The F1 score combines precision and recall through their harmonic mean. 
Our dataset has imbalanced class distribution {(82\% non-exploitation; 18\% exploitation)}; thus, we focus on F1 score here.
Due to space constraints, we do not include precision and recall results that show similar   trends.

%


\begin{table} \vspace{0.05in}
\caption{Temporal classification accuracy (F1 scores): different features and classifiers using equal window temporal splits.}
\centering
\label{table: impact_of_approach}
\scalebox{0.83}{
\begin{tabular}{|l|r|r|r|r|r|r|r|r|r|r|} 
\hline
\multirow{3}{*}{Split ID} & \multicolumn{10}{c|}{Approach}                                                                                                                                                                                                                                                                                                                                                                                                                                                                                                                                                                                                                                                                                           \\ 
\cline{2-11}
                          & \multicolumn{3}{c|}{Doc2Vec}                                                                                                                                                                                                         & \multicolumn{3}{c|}{TI2Vec}                                                                                                                                                                & \multicolumn{3}{c|}{BERT Embeddings}                                                                                                                                                                                                 & \multicolumn{1}{l|}{\multirow{2}{*}{TIBERT}}  \\ 
\cline{2-10}
                          & \multicolumn{1}{l|}{\begin{tabular}[c]{@{}l@{}}Naive \\Bayes\end{tabular}} & \multicolumn{1}{l|}{\begin{tabular}[c]{@{}l@{}}Decision \\Tree\end{tabular}} & \multicolumn{1}{l|}{\begin{tabular}[c]{@{}l@{}}Ada\\Boost\end{tabular}} & \multicolumn{1}{l|}{\begin{tabular}[c]{@{}l@{}}Naive \\Bayes\end{tabular}} & \multicolumn{1}{l|}{\begin{tabular}[c]{@{}l@{}}Decision \\Tree\end{tabular}} & \multicolumn{1}{l|}{AdaBoost} & \multicolumn{1}{l|}{\begin{tabular}[c]{@{}l@{}}Naive \\Bayes\end{tabular}} & \multicolumn{1}{l|}{\begin{tabular}[c]{@{}l@{}}Decision \\Tree\end{tabular}} & \multicolumn{1}{l|}{\begin{tabular}[c]{@{}l@{}}Ada\\Boost\end{tabular}} & \multicolumn{1}{l|}{}                         \\ 
\hline
2                         & 0.93                                                                        & 0.64                                                                         & 0.86                                                                    & 0.9                                                                         & 0.75                                                                         & 0.83                          & 0.68                                                                        & 0.63                                                                         & 0.75                                                                    & 0.72                                          \\ 
\hline
3                         & 0.87                                                                        & 0.81                                                                         & 0.72                                                                    & 0.82                                                                        & 0.87                                                                         & 0.83                          & 0.8                                                                         & 0.75                                                                         & 0.78                                                                    & 0.92                                          \\ 
\hline
4                         & 0.69                                                                        & 0.50                                                                         & 0.55                                                                    & 0.37                                                                        & 0.68                                                                         & 0.74                          & 0.36                                                                        & 0.44                                                                         & 0.33                                                                    & 0.83                                          \\ 
\hline
5                         & 0.06                                                                        & 0.55                                                                         & 0.75                                                                    & 0.5                                                                         & 0.81                                                                         & 0.83                          & 0.79                                                                        & 0.75                                                                         & 0.83                                                                    & 0.89                                          \\ 
\hline
6                         & 0.57                                                                        & 0.47                                                                         & 0.64                                                                    & 0.6                                                                         & 0.63                                                                         & 0.65                          & 0.42                                                                        & 0.49                                                                         & 0.57                                                                    & 0.69                                          \\ 
\hline
7                         & 0.49                                                                        & 0.46                                                                         & 0.67                                                                    & 0.59                                                                        & 0.6                                                                          & 0.67                          & 0.67                                                                        & 0.58                                                                         & 0.63                                                                    & 0.66                                          \\ 
\hline
8                         & 0.60                                                                        & 0.60                                                                         & 0.74                                                                    & 0.49                                                                        & 0.7                                                                          & 0.75                          & 0.58                                                                        & 0.67                                                                         & 0.75                                                                    & 0.78                                          \\ 
\hline
9                         & 0.32                                                                        & 0.40                                                                         & 0.64                                                                    & 0.53                                                                        & 0.61                                                                         & 0.65                          & 0.57                                                                        & 0.63                                                                         & 0.61                                                                    & 0.65                                          \\ 
\hline
10                        & 0.51                                                                        & 0.76                                                                         & 0.74                                                                    & 0.46                                                                        & 0.49                                                                         & 0.68                          & 0.54                                                                        & 0.54                                                                         & 0.74                                                                    & 0.84                                          \\ 
\hline
Average                   & 0.56                                                                        & 0.58                                                                         & 0.70                                                                    & 0.58                                                                        & 0.68                                                                         & 0.74                          & 0.60                                                                        & 0.61                                                                         & 0.67                                                                    & 0.78                                          \\
\hline
\end{tabular}}
\arrayrulecolor{black}
\vspace{-0.2in}
\end{table}

%
%

%
%
%
%

\textbf{Impact of feature extraction. }
Table \ref{table: impact_of_approach} summarizes the classification results for different feature extraction approaches using the equal window split. 
We break down the results for 4 different feature extraction approaches: Doc2Vec, TI2Vec, BERT, and TIBERT. 
The best average classification accuracy of 78\% is achieved by TIBERT, which  
  turns out to be a better approach than the pre-trained BERT embeddings as well as Doc2Vec and TI2Vec. 
TIBERT achieves 4{\%} better F1 score than TI2Vec.
As compared to TI2Vec, TIBERT  produces word representations that are context-dependent. 
Thus, the same word would have multiple embeddings differing by the context of the words around them whereas TI2Vec is a context-independent embedding. 
%
%
Overall, TI2Vec and TIBERT significantly outperform their vanilla pre-trained Doc2Vec and BERT counterparts. 
The improvement in accuracy can be attributed to the specialization of these embeddings on a large corpus of IoC data from many different TI feeds. 
By doing so, TI2Vec and TIBERT capture the context of security events in TI feeds, being able to better characterize its similarity to other security events when compared to   pre-trained embeddings.

\textbf{Impact of classifier. }
Comparing different machine learning classifiers\footnote{except for TIBERT which has a built-in classifier}, we note that AdaBoost outperforms Naive Bayes and Decision Tree. 
Different classifiers appear to handle the precision-recall trade-off differently. 
For instance, not shown in Table \ref{table: impact_of_approach}, the Decision Tree classifier is inclined to optimize for precision while the Naive Bayes classifier is inclined to optimize for recall. 
AdaBoost achieves the best balance between precision and recall. 
AdaBoost outperforms Decision Tree by 12\% for Doc2Vec, 6\% for TI2Vec, and 6\% for BERT in terms of overall F1 score.

\begin{table}
\vspace{0.05in}
\caption{TIBERT temporal classification accuracy (F1 scores)}
\label{table: temporal_bert_results}
\footnotesize
\resizebox{\columnwidth}{!}{
\begin{tabular}{|l|l|l|l|l|l|l|l|l|l|l|} 
\hline
\multirow{2}{*}{Split Type} & \multicolumn{9}{c|}{Split ID}                                & \multirow{2}{*}{Average}   \\ 
\cline{2-10}
                            & 2    & 3    & 4    & 5    & 6    & 7    & 8    & 9    & 10   &                            \\ 
\hline \hline
Equal Window                & 0.72 & 0.92 & 0.83 & 0.89 & 0.69 & 0.66 & 0.78 & 0.65 & 0.84 & \multicolumn{1}{c|}{0.78}  \\ 
\hline
Aggregate Window                 & NaN  & 0.86 & 0.37 & 0.17 & 0.56 & 0.77 & 0.77 & 0.67 & 0.84 & \multicolumn{1}{c|}{0.63}  \\
\hline
\end{tabular}}
\arrayrulecolor{black}\vspace{-0.1in}
\end{table}

\textbf{Impact of temporal splits. }
Table \ref{table: temporal_bert_results} reports F1 score for TIBERT for two types of temporal splits: equal and aggregate window, where the latter 
%
 accumulates training data over time. 
However, we note that the classification accuracy for aggregate window splits does not monotonically increase across splits -- it increases initially and then decreases sharply for split 4 and then improves again eventually as the size of the train set increases. 
The equal window split type explores the impact of equal-sized train and test sets. 
Note that the test sets are comparable between both split types, but aggregate window uses more training data.  
%
%
For split  4, the sharp drop in F1 scores is explained by the large difference in the vocabulary between train and test sets.\footnote{For split 4, the difference in vocabulary between the test and train sets was about 44,000 words. The difference was determined by computing the number of words in the test set that did not appear in the train set.}

Comparing aggregate window and equal window, we note that equal window achieves better classification performance.
For each split, limiting our training data to only recent events instead of aggregating it generally resulted in a higher F1 score. 
For instance, the equal window split average (78\%) outperformed  aggregate window split average (63\%) for TIBERT. 
Thus, we conclude that the limiting the training data to the most recent ultimately helps the classifier to more accurately detect exploitation events than using more training data.


\begin{table}
\caption{TIBERT spatial classification accuracy (F1 scores) } 
\label{table: spatial_results}
\footnotesize
\centering
\resizebox{\columnwidth}{!}{
\begin{tabular}{|l|l|l|l|l|l|l|l|} 
\hline
\multirow{2}{*}{Split Type} & \multicolumn{6}{c|}{Feed Omitted}                                                                                                                                                                    \\ 
\cline{2-7}
                            & \CERTBund  & \CIRCL     & \inThreat  & \CthulhuSPRL & \DCSO      &  \INCIBE     \\ 
\hline
Aggregate Window                & \multicolumn{1}{c|}{0.45} & \multicolumn{1}{c|}{0.79} & \multicolumn{1}{c|}{0.94} & \multicolumn{1}{c|}{0.92}   & \multicolumn{1}{c|}{0.89}  & \multicolumn{1}{c|}{0.60}  \\
\hline 
\end{tabular}}\vspace{-0.2in}
\end{table}

\textbf{Impact of TI feed. }
Next, we report the accuracy results for joint temporal+spatial splits. 
These   splits allow us evaluate the transferability of the classification model to an unseen TI feed. 
Table \ref{table: spatial_results} shows the F1 scores for the various temporal+spatial splits, grouped by the feed omitted from the train set (which is the only feed present in the test set).
Since valid F1 scores depend on the presence of exploit events in the test set, we further restrict analysis to temporal+spatial splits for feeds that contain at least one exploit event in the test set.
We focus our analysis on the TIBERT aggregate window approach as it yielded the highest F1 scores.
We observe a strong correlation between the findings of Figure \ref{figure:sankey} and Table~\ref{table: spatial_results}. 
We note that the feeds with IoCs that are mostly ``sinks" also have higher average F1 scores when that feed is omitted in the temporal+spatial splits.
%
%
For example, Table \ref{table: spatial_results} shows that the omission of {\DCSO} yields a high F1 score (89\%). 
{\DCSO} also notably has the most number of Hash and IP IoCs (2088) that are sinks of IoCs from other feeds in the dataset. 
Likewise, the omission of {\CIRCL}, which has the second highest number of sink Hash and IP IoCs (1095), yields F1 scores that are in the top 4 highest scores for aggregate window split.

In the same vein, the omission of {\CERTBund} yields a relatively low F1 score (45\%). While   {\CERTBund} has high IoC overlap with other feeds,  
Figure \ref{figure:sankey} shows that the majority ($\approx$ 85\%) of \CERTBund's IoC overlap occurs when it is a source and not a sink. 
Exceptions to these general findings do exist, however. 
For example, as Figure~\ref{figure:sankey} shows, {\CthulhuSPRL}  is more often a source than a sink, yet its omission yields a high F1 score (92\%). 
This may indicate that {\CthulhuSPRL}  shares many similar IoCs with other feeds, despite not having an abundance of exact IoC overlap.

Overall, these findings suggest that our model's ability to detect exploitation events in previously unseen TI feeds depends on how the IoCs in the training set relate to those in the testing set. 
%
High accuracy exploitation detection is possible under spatial restrictions if the training data has same/similar IoCs as those present in the testing feed. 

\begin{figure}
    \centering
    \begin{subfigure}[b]{0.47\columnwidth}
        \includegraphics[width=1\columnwidth]{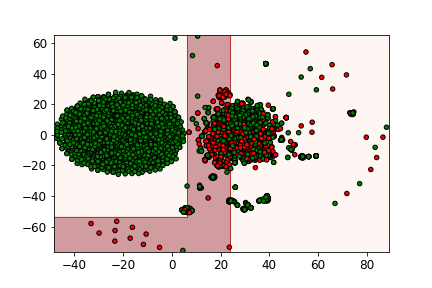}
        \caption{Doc2Vec}
     \end{subfigure}
     \begin{subfigure}[b]{0.47\columnwidth}
      \includegraphics[width=1\columnwidth]{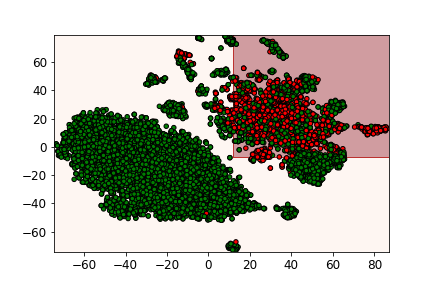}
      \caption{TI2Vec}
    \end{subfigure}
     \caption{Embeddings   input into a decision tree classifier} 
     \label{fig:bertdoc2vec}\vspace{-0.25in}
\end{figure}

\textbf{TIBERT feature ablation analysis. } 
We conduct feature ablation analysis to assess the relative contribution of different IoC categories.
The classification accuracy, as measured using the F1 score,  for top three IoC categories, 
 equals 40\%, 39\% and 66\% for network activity, payload delivery, and external analysis, respectively. Across all categories, accuracy equals 78\%. 
We note that classification accuracy for network activity IoCs is better than both payload delivery and external analysis. 
The results are inline with the earlier finding that our TI feed dataset contains the most network activity IoCs (557,532), followed by payload delivery (220,321), and then external analysis (38,531). 
Overall, all IoCs of different categories instead of just one, such as network activity, yields the best classification accuracy (78\% vs 66\%).
Thus, we conclude that while some IoC categories are more helpful than others, they complement each other and using them together outperforms standalone classification performance. 

  \textbf{Visualization. } 
  \label{sec:visual}
Figure~\ref{fig:bertdoc2vec} illustrates  Doc2Vec and TI2Vec embeddings.  
Using TSNE, we  visualize how the embeddings produced by Doc2Vec and TI2Vec spread over the plane.  
Each point corresponds to an event.  
Green and red points  correspond to non-exploitation and exploitation events, respectively.  
Whereas both Doc2Vec and TI2Vec produced two clearly distinguishable clusters,   clusters produced by TI2Vec are  easier to separate using a simple decision tree.
%
%
A simple decision tree, with two levels, is   able to classify the events with a 55\% and 66\% F1-score using Doc2Vec and TI2Vec, respectively.  
\ifthenelse{\boolean{compact8pages}}{For further details, we refer to~\cite{techrep}.}{ 
Those numbers are in agreement with  Table~\ref{table: impact_of_approach}, noting that the average F1-scores of 58\% and 68\% reported in Table~\ref{table: impact_of_approach} {for Doc2Vec and TI2Vec were obtained with  deeper decision trees and the original embeddings. 
As expected, using shallow  trees and  simplified TSNE embeddings    caused a slight decrease in performance. 
} }

\ifthenelse{\boolean{compact8pages}}{}{    
\textbf{Applications of TIBERT. }
The exploitation events predicted by TIBERT can be utilized to update the temporal exploitability score of CVSS/EPSS related to the availability of PoC  or fully functional exploits. 
We focus  on temporal metrics since TIBERT generates predictions based on the past history of a event. The \textit{Exploit Code Maturity} metric describes the maturity an exploit. The output of TIBERT can be used to automatically revise this metric, such as updating it from ``Proof of concept code'' to ``Functional exploit exists''.   

TIBERT analyzes large volumes of loosely structured TI feeds to   enhance the quality of CVSS/EPSS scores, saving time and effort for  experts. Using TIBERT, experts can  also prioritize  data from catalogs such as   CISA KEV~\cite{CISAKEV} and Google Project Zero~\cite{GoogleProjectZero} based on exploitation events relevant to their organizations, as reflected by representative TI feeds.

}

\section{Related Work}
\label{sec: background}

 
There is growing interest  
to better understand the value of information in TI feeds. 
Early work focused on TI feeds   targeting specific threat niches \cite{Sheng09phishingblacklists,Pitsillidis12spamfeeds}.
%
%
%
Phishing and spamming, for instance,  are classical themes in the realm of TI feeds~\cite{Sheng09phishingblacklists, Pitsillidis12spamfeeds,liao2016acing,bouwman2020different}.
Sheng et al. \cite{Sheng09phishingblacklists} analyzed the effectiveness of 8 different phishing URL blacklists. 
They found that the quality of different blacklists varies with respect to coverage and information freshness.
%
%
More recently, Li et al.~\cite{li19tealeaves} analyzed 47 different TI feeds, 
also pointing that 
%
the  quality of information across   TI feeds varies substantially.
In conclusion, the main takeaway from prior work on empirical analysis of TI feeds is that they vary widely in terms of the quantity, quality and coverage\ifthenelse{\boolean{compact8pages}}{.}{ \cite{tounsi2018survey}.}


%

%
%

Our work builds on prior research in this space by analyzing 191 different general purpose TI feeds. 
However, instead of a similar comparative analysis of different TI feeds, our main goal is to develop an automated approach to identify vulnerability exploitation events from TI feeds.  
%

Different sources of information can be used for predicting various states in a vulnerability's lifecycle, such as discovery, weaponization, exploitation, reporting and patching~\ifthenelse{\boolean{compact8pages}}{\cite{Almukaynizi17identificationexploits,Wang17patchingICs,Almukaynizi19predictexploit}.}{\cite{householder2020historical,yin2022vulnerability,Sabottke15exploitsocial,Almukaynizi17identificationexploits,Wang17patchingICs,Almukaynizi19predictexploit}.}
{Models for vulnerability lifecycles  are instrumental to parametrize CVSS/EPSS~\ifthenelse{\boolean{compact8pages}}{\cite{fuhrmann2022cyber}.}{\cite{varela2022feature,fuhrmann2022cyber}.} However, whereas  CVSS/EPSS are vulnerability-oriented, requiring a CVE ID to assess its exploitability/risk\ifthenelse{\boolean{compact8pages}}{,}{ \cite{shahid2021cvss},} our work is TI event-oriented. Our work (1) aims at determining if a given event corresponds to an exploitation   and (2) is event-driven, leveraging the description of security incidents for that matter. 
TIBERT's detection of exploitation   in the wild essentially provides the ground truth for EPSS scores,  complementing catalogs such as  CISA KEV~\cite{CISAKEV} and Google Project Zero~\cite{GoogleProjectZero}. 
} 

\ifthenelse{\boolean{compact8pages}}{}{
Sabottke et al. \cite{Sabottke15exploitsocial} and Suciu et al.~\cite{suciu2021expected} showed that  social media data can be leveraged to detect vulnerability exploits in the wild. 
They trained a machine learning classifier using keyword and other metadata features to detect proof-of-concept and real-world exploits. 
Complementary to those works,~\cite{moreno2023cream,Almukaynizi17identificationexploits,Almukaynizi19predictexploit} 
showed that data from darknet web forums and other traditional sources can be combined to detect vulnerability exploitation.
They trained various machine learning classifiers to predict the likelihood of vulnerability exploitation.
%
%
%
%


Our work is similar in spirit to this line of research on training  classifiers to predict different stages of a vulnerability's lifecycle.
Differently from prior work, we focus on training machine learning models to specifically detect instances of vulnerability exploitation from  loosely structured TI feeds.

}

\section{Conclusion \& Future Work}
\label{sec:conclusion}

We proposed a machine learning pipeline for automated analysis of loosely structured information in TI feeds. 
Then, aiming towards detecting vulnerability exploitation events from TI feeds, we proposed  TIBERT, that can be used to  automatically update CVSS/EPSS scores and to prioritize data from CISA KEV and Google Project Zero catalogs, by leveraging rich temporal data from TI feeds. This paves the way toward contextualizing data from those catalogs 
with minimal effort from security experts~\cite{jacobs2021exploit}.
We envision that our work opens up a number of  directions for future research.  In particular,  
%
%
there is opportunity to train the considered  embeddings on a larger corpus   to improve   coverage and diversity of vocabulary.  Our machine learning pipeline can also be readily adapted to detect other relevant vulnerability lifecycle events, e.g., related to  increase in    weaponization maturity.
\bibliographystyle{IEEEtran}
\bibliography{main}
%

\end{document}